\documentclass{ws-ijmpd}

\newcommand{\be}{\begin{equation}}
\newcommand{\ee}{\end{equation}}
\newcommand{\ba}{\begin{eqnarray}}
\newcommand{\ea}{\end{eqnarray}}

\newcommand{\n}[1]{\label{#1}}
\newcommand{\eq}[1]{(\ref{#1})}

\newcommand{\hhh}{\, ,\hspace{0.2cm}}
\newcommand{\tti}{\tilde{t}}
\newcommand{\tr}{\tilde{r}}
\newcommand{\tth}{\tilde{\theta}}
\newcommand{\tp}{\tilde{\phi}}
\newcommand{\tm}{\tilde{\mu}}
\newcommand{\tn}{\tilde{\nu}}
\newcommand{\ce}{{\cal E}}
\newcommand{\cl}{{\cal L}}
\newcommand{\ck}{{\cal K}}

\begin{document}

\markboth{A. M. AL ZAHRANI, VALERI P. FROLOV, and ANDREY A. SHOOM}
{PARTICLE DYNAMICS IN WEAKLY CHARGED EXTREME KERR THROAT}

%
\catchline{}{}{}{}{}
%

\title{PARTICLE DYNAMICS IN WEAKLY CHARGED\\
EXTREME KERR THROAT}

\author{A. M. AL ZAHRANI}

\address{Theoretical Physics Institute, University of Alberta,\\
Edmonton, Alberta, T6G 2G7, Canada\\
ama3@ualberta.ca}

\author{VALERI P. FROLOV}

\address{Theoretical Physics Institute, University of Alberta,\\
Edmonton, Alberta, T6G 2G7, Canada\\
vfrolov@ualberta.ca}

\author{ANDREY A. SHOOM}

\address{Theoretical Physics Institute, University of Alberta,\\
Edmonton, Alberta, T6G 2G7, Canada\\
ashoom@ualberta.ca}

\maketitle

\begin{history}
\received{Day Month Year}
\revised{Day Month Year}
\comby{Managing Editor}
\end{history}

\begin{abstract}

We study dynamics of a test charged particle moving in a weakly charged
extreme Kerr throat. Dynamical equations of the particle motion are
solved in quadratures. We show explicitly that the Killing tensor of
the Kerr spacetime becomes reducible in the extreme Kerr throat
geometry. Special types of motion of particles and light are
discussed.

\end{abstract}


\section{Introduction}	

\noindent In this paper, we study  motion of charged particles and light in a
throat of the extreme Kerr metric. Such a model is an idealization. However,
it allows one to simplify many calculations. On the other hand, it gives
better understanding of the near horizon physics for rapidly rotating
black holes. The motivation for this study is the following.
One expects that most of the astrophysical black holes are rapidly
rotating. Even if the angular momentum of a collapsing progenitor  star
is small, the falling matter of the accretion disk can spin up
the formed black hole.\cite{bardeen,lynd} If the mass of the
accretion disk is large enough, such a black hole of mass $M$
may become practically extreme, with the angular momentum
$J=Ma\approx M^2$. The radiation from the disk slightly restricts the
limiting value of the rotation parameter, so that
$a/M\approx 0.998$.\cite{thorne} The recent astrophysical
observations confirm that at least some of the black holes are rapidly
rotating.\cite{clin}

\noindent The geometry of the extreme Kerr throat was recently
discussed in connection with the Kerr/CFT correspondence (see, e.g.,
Ref.~\refcite{bred} and references there in). This approach is motivated by
the fact that the limiting geometry of the throat of the extreme Kerr
metric generates a non-singular vacuum solution with an enhanced
symmetry group $SL_2(R) \times U(1)$, which is similar to
$AdS_2\times S^2$ metric. To obtain this limiting geometry one uses
the procedure proposed by Bardeen\cite{bard} and developed in the
paper Ref.~\refcite{EK}.

\noindent One can generalize the consideration to the case of extreme
black holes in the Einstein-Maxwell theory.  For example, one can use
the exact solution for a rotating black hole with electric charge
(the Kerr-Newman metric\cite{KN}) and for a rotating black hole in the
external homogenous field (the Melvin solution\cite{Mel1,Mel2}). In this paper,
in order to include electromagnetic effects we use a weak field
approximation. This approximation allows one to simplify calculations
considerably. At the same time, it can be validated in many applications.
The charge-to-mass ratios for particles are usually very large, so that the test electromagnetic
field may affect their motion even if its gravitational effect on the
background geometry is weak (see, e.g., Refs.~\refcite{AG,Aliev,FrSh}).

\noindent This paper is organized as follows: In Section~2 we discuss a
weakly charged and magnetized black hole and the reduction of
its geometry and fields in its throat in the extreme Kerr limit. We demonstrate that, as a
result of the reduction, the original Killing tensor of the Kerr
spacetime becomes reducible. Equations of motion for charged
particles and light in the extreme Kerr throat are derived in
Section~3. We study the radial equation of motion in Section~4. The angular
equation of motion for some special cases is discussed in Sections~5 and 6.
Section 7 contains a brief summary. In this paper we
use the sign conventions adopted in Ref.~\refcite{MTW} and units where
$G=c=1$.

\section{Weakly Charged Extreme Kerr Throat}

\subsection{Kerr black hole in weak electromagnetic field}

The Kerr metric in the Boyer-Linquist coordinates
$\tilde{x}^{\tm}=(\tti,\tr,\tth,\tp)$ reads
\be\n{K}
d\tilde{s}^{2}=\tilde{g}_{\tm\tn}d\tilde{x}^{\tm}d\tilde{x}^{\tn}
=-\Sigma\frac{\Delta}{A}d\tti^{\,2}
+\frac{\Sigma}{\Delta}d\tr^{2}+\Sigma d\tth^{2}
+\frac{A}{\Sigma}\left(d\tp-\frac{2aM\tr}{A}d\tti\right)^{2}\sin^{2}\tth\,,
\ee
where
\be
\Sigma=\tr^{2}+a^{2}\cos^{2}\tth\hhh \Delta=
\tr^{2}-2M\tr+a^{2}\hhh A=(\tr^{2}+a^{2})^{2}-a^{2}\Delta\sin^{2}\tth\,.
\ee
This metric describes a Kerr black hole of the mass $M$ and the
angular momentum $J=aM$, where $0\leq|a|\leq M$. The Kerr metric possesses two commuting Killing vectors
\be\n{kv}
\xi^{\tm}_{(\tti)}=\delta^{\tm}_{\tti}\hhh \xi^{\tm}_{(\tp)}
=\delta^{\tm}_{\tp}\,,
\ee
and the Killing tensor
\be\n{kt}
\tilde{K}^{\tm\tn}=a^{2}\sin^{2}\tth\,\xi^{\tm}_{(\tti)}\xi^{\tn}_{(\tti)}
+2a\,\xi^{\tm}_{((\tti)}\xi^{\tn}_{(\tp))}+
\frac{1}{\sin^{2}\tth}\xi^{\tm}_{(\tp)}\xi^{\tn}_{(\tp)}+
\delta^{\tm}_{\tth}\delta^{\tn}_{\tth}-a^{2}(1+\cos^{2}\tth)
\tilde{g}^{\tm\tn}\,.
\ee
The Killing tensor is presented in a form that is more convenient for our purposes,
which differs from the standard one, corresponding to the Carter
constant, by an additional term $-a^{2}\tilde{g}^{\tm\tn}$.

In a Ricci flat spacetime, a Killing vector obeys the same equations as
an electromagnetic potential for the source-free electromagnetic field in the Lorentz gauge.\cite{Wald}
Using the Killing vectors in the Kerr spacetime, we have
\be\n{elp}
A^{\tm}= \left(aB-\frac{Q}{2M}\right) \xi^{\tm}_{(\tti)}
+ \frac{B}{2}\xi^{\tm}_{(\tp)}\, .
\ee For such a choice, the parameter $B$ represents the strength of the test magnetic field,
which is uniform and homogeneous at  the spatial infinity, where it
is directed along the axis of rotation of the black hole. The parameter $Q$ is the
electric charge of the black hole,
\be
Q={1\over 4\pi}\int_{\sigma} F^{\mu\nu}d\sigma_{\mu\nu}\, .
\ee
Here $\sigma$ is a 2D surface surrounding the black hole.
We assume that the electric and magnetic fields are weak in the sense
that their effect on the spacetime curvature near the black hole
horizon $\tilde{r}=r_{+}=M+\sqrt{M^{2}-a^{2}}$
is negligible, i.e., the following conditions hold
\be
|Q|\ll r_{+}\hhh |B|\ll 1/r_{+} \,.
\ee

It was shown by Carter\cite{Cart:BOOK} that a rotating black
hole immersed into a homogeneous magnetic field acquires a
non-vanishing chemical potential. As a result, in the presence of free
charged particles, their accretion will charge the black hole until
the condition $Q=2aMB$ is satisfied. Another effect, which might
generate a non-vanishing charge of the black hole, is the difference in
the accretion rate for electrons and protons of the surrounding
plasma in the presence of radiation from the accreting matter. The
ratio of the electrostatic force to the gravitational one for a
particle of mass $m$ and charge $q$ in the field of a charged black hole is
$qQ/mM$. One can expect that $Q/M$ cannot be significantly greater
than $m/q$ (see, e.g., Ref.~\refcite{Wald}). We shall not specify the value
of $Q$ in our work and keep it as a free parameter.

\subsection{Weakly charged extreme Kerr throat}

The extreme Kerr black hole corresponds to $|a|=M$. Consider the following
coordinate transformation:\cite{bard}
\be\n{tr}
\tti=\tau M/\lambda\hhh \tr=M(1+\lambda\rho)\hhh \tth
=\theta\hhh\tp=\phi+\tau/(2\lambda)\,.
\ee
The limit $\lambda\to0$ gives the geometry of the extreme Kerr throat,\cite{EK}
\ba\n{ke}
d\sigma^{2}&=&(1+\cos^{2}\theta)\left[-\frac{\rho^{2}}{4}d\tau^{2}
+\frac{d\rho^{2}}{\rho^{2}}+d\theta^{2}\right]+
\frac{4\sin^{2}\theta}{1+\cos^{2}\theta}\left(d\phi+\frac{\rho}{2}d\tau\right)^{2}\,,
\ea
where $d\sigma^{2}=M^{-2}d\tilde{s}^{2}$. The throat horizon is at
$\rho=0$. The limiting metric (\ref{ke})
has four Killing vectors,\cite{EK}
\be\n{kve}
\xi^{\mu}_{(1)}=\delta^{\mu}_{\tau}\hhh \xi^{\mu}_{(2)}
=\delta^{\mu}_{\phi}\hhh \xi^{\mu}_{(3)}=\tau\delta^{\mu}_{\tau}
-\rho\,\delta^{\mu}_{\rho}\hhh \xi^{\mu}_{(4)}=\left[\frac{\tau^{2}}{2}
+\frac{2}{\rho^{2}}\right]\delta^{\mu}_{\tau}-\tau\rho\,\delta^{\mu}_{\rho}
-\frac{2}{\rho}\delta^{\mu}_{\phi}\,.
\ee
The symmetry group generated by these Killing vectors is $SL_2(R)\times U(1)$.

In the same limit, the 4-potential \eq{elp} takes the form
\be\n{vecp}
A^{\mu}=\frac{Q}{4M} \xi^{\mu}_{(2)}\, .
\ee
It is easy to check that
\be
[\xi_{(2)},\xi_{(i)}]=0 \hhh i=1,2,3,4\,.
\ee
It means that the vector-potential \eq{vecp} remains constant along the orbits of the Killing vectors,
\be\n{Asym}
L_{\xi_{(i)}}A=0\hhh i=1,2,3,4\,.
\ee
The form of the 4-potential implies that in the extreme Kerr throat
only the field produced by the charge $Q$ survives. This is a manifestation
of the well known property of rapidly rotating black holes: The flux
lines of the axisymmetric magnetic field are expelled in the extreme limit and the
black hole behaves as a perfect diamagnetic.\cite{BiDv} In this sense,
one can see the emergence of the Meissner effect in the extremely
rotating black holes (as well as branes).\cite{Gibb}

\subsection{Reduction of the Killing tensor}

As a result of the limiting process, the symmetries of the extreme
Kerr throat metric are enhanced. In particular, the symmetry which is
hidden in the Kerr spacetime becomes an explicit spacetime symmetry.
Applying the transformation \eq{tr} and taking the limit
$\lambda\to0$ of the Killing tensor \eq{kt} we derive
\be\n{rkt1}
K^{\mu\nu}=\frac{4}{\rho^2}\delta^{\mu}_{\tau}\delta^{\nu}_{\tau}
-\rho^{2}\delta^{\mu}_{\rho}\delta^{\nu}_{\rho}
-\frac{4}{\rho}\delta^{(\mu}_{\tau}\delta^{\nu)}_{\phi}
+\delta^{\mu}_{\phi}\delta^{\nu}_{\phi}\,.
\ee
Using the Killing vectors \eq{kve} we can present this Killing tensor
as follows:
\be\n{rkt2}
K^{\mu\nu}=2\xi^{(\mu}_{(1)}\xi^{\nu)}_{(4)}-
\xi^{\mu}_{(3)}\xi^{\nu}_{(3)}+\xi^{\mu}_{(2)}\xi^{\nu}_{(2)}\,.
\ee
Thus, the Killing tensor \eq{kt}, responsible for the hidden
symmetries, becomes reducible in the extreme
Kerr throat geometry (see also Refs.~\refcite{Anton,Ras} ).

Using the commutators for the Killing vectors \eq{kve},
one can show that
\be\n{sym}
[{\xi}_{(1)},\xi_{(2)}]=0\hhh L_{\xi_{(1)}}K=L_{\xi_{(2)}}K=0\,.
\ee

\section{Equations of Motion}

In this section we analyze dynamics of both massive charged particles and
light (photons) moving in a weakly charged extreme Kerr throat.
To deal with these two cases simultaneously, we write the dynamical
equations in the form
\be\n{d}
\dot{u}^{\mu}=\varepsilon f^{\mu}\, .
\ee
For the light motion $\varepsilon=0$, so that the right hand side
vanishes. In this case the `dot' means the (covariant) derivative with
respect to an affine parameter $\ell$ and
$u^{\mu}=dx^{\mu}/d\ell$. For a massive particle with mass $m$ and
charge $q$, the `dot'
denotes the (covariant) derivative with respect to the proper time
$\sigma$, $u^{\mu}=dx^{\mu}/d\sigma$ and
\be
\varepsilon\equiv u^{\mu}u_{\mu}=-1\, .
\ee
We also denote
\be
f^{\mu}=-{q\over m}F^{\mu}_{\,\,\,\,\nu}u^{\nu}\,.
\ee
The electromagnetic field tensor $F_{\mu\nu}=A_{\nu,\mu}-A_{\mu,\nu}$
is calculated for the 4-potential \eq{vecp}.

The commuting set of the symmetry generators $(\xi_{(1)},\xi_{(2)},K)$
can be used to construct the following integrals of motion\footnote{
For the electromagnetic potential obeying the condition \eq{Asym}, the
quantities $\xi^{\mu}_{(3)}{\cal P}_{\mu}$ and $\xi^{\mu}_{(3)}{\cal
P}_{\mu}$ are also conserved. For our purpose, it is sufficient to use
instead of them the integral of motion ${\cal K}$, which is in involution with ${\cal E}$ and ${\cal
L}$.}:

\be\n{intm}
\ce\equiv-\xi^{\mu}_{(1)}{\cal P}_{\mu}\hhh
\cl\equiv\xi^{\mu}_{(2)}{\cal P}_{\mu}\hhh
\ck\equiv K^{\mu\nu}{\cal P}_{\mu}{\cal P}_{\nu}-\cl^{2}\,.
\ee
Here
\be
{\cal P}^{\mu}=u^{\mu}-\frac{\varepsilon q}{m}A^{\mu}=
(\dot{\tau},\dot{\rho},\dot{\theta},\dot{\phi}
-\varepsilon b)\hhh b=\frac{qQ}{4mM}\,.
\ee
For a massive particle, ${\cal P}^{\mu}$ is the generalized  specific
momentum per unit mass, whereas for a massless particle
${\cal P}^{\mu}=u^{\mu}$.
The first constant of motion $\ce$
corresponds to energy, the second one $\cl$ corresponds to
the generalized azimuthal angular momentum, and the third one $\ck$ is
analogous to the Carter constant.
The integrals of motion \eq{intm} are in involution. Let us emphasize that the integrals of motion ${\cal E}$
and ${\cal L}$ are the `throat' analogues of the energy
$E=-\xi^{\mu}_{{(\tau)}} {\cal P}_{\mu}$ and the generalized angular momentum
$L=\xi^{\mu}_{{(\phi)}} {\cal P}_{\mu}$ of the original Kerr
spacetime. Namely, applying the transformation (\ref{tr}) to $E$ and $L$ we derive
\be\n{EKEL}
\tilde{E}=\lambda E(\lambda)+
\frac{1}{2}L(\lambda)\hhh \tilde{L}=L(\lambda)\,,
\ee
where
\be\n{limEL}
\lim_{\lambda\rightarrow0}E(\lambda)=\ce \hhh
\lim_{\lambda\rightarrow0}L(\lambda)=\cl\,.
\ee

Using the conserved quantities and
the 4-velocity normalization we derive the system of the first order equations for the
particle motion in a weakly charged extreme Kerr throat,
\ba
\dot{\tau}&=&\frac{2(\cl\rho+2\ce)}{\rho^{2}(1+\cos^{2}\theta)}\,,\n{t}\\
\dot{\rho}&=&\pm\frac{(4\ce^{2}+4\ce\cl\rho-\ck\rho^{2})^{1/2}}{1+\cos^{2}\theta}\,,\n{r}\\
\dot{\theta}&=&\pm\frac{(A(1+\cos^{2}\theta)\sin^{2}\theta+2B\sin^{2}\theta-4\cl^{2})^{1/2}}{2\sin\theta(1+\cos^{2}\theta)}\,,\n{th}\\
\dot{\phi}&=&\frac{\cl\rho(\cos^{4}\theta+6\cos^{2}\theta-3)-8\ce\sin^{2}\theta}{4\rho\sin^{2}\theta(1+\cos^{2}\theta)}+\varepsilon b\,,\n{ph}
\ea
where
\be
A=(\cl-4\varepsilon b)^{2}+4\varepsilon\hhh B=3\cl^{2}+2\ck-16\varepsilon^{2}b^{2}\,.\n{A}
\ee
According to the Liouville theorem, this system of equations is completely integrable. Thus, the problem of the particle motion can be reduced to one of quadratures. In particular, for the particle motion in the $(\rho,\theta)$ sector one has
\be\n{rth}
\int_{\rho_{0}}^{\rho}\frac{d\rho'}{\sqrt{4\ce^{2}+4\ce\cl\rho'-\ck\rho'^{2}}}=2\int_{z_{0}}^{z}\frac{dz'}{\sqrt{A(1-z'^{4})+2B(1-z'^{2})-4\cl^{2}}}\,,
\ee
where $\rho_{0}$ and $z_{0}$ are initial values of $\rho$ and $z$, $z=\cos\theta$. The following conditions must hold:
\ba
4\ce^{2}+4\ce\cl\rho-\ck\rho^{2}\geq0\,,\n{c1}\\
A(1-z^{4})+2B(1-z^{2})-4\cl^{2}\geq0\,,\n{c2}\\
\cl\rho+2\ce>0\,.\n{c3}
\ea
Here the last condition means that the 4-velocity vector $u^{\mu}$ is future directed, or, what is the same, local energy $\ce_{\text{loc}}$ of the particle is positive:
\be\n{loce}
\ce_{\text{loc}}=\frac{\dot{\tau}\rho}{2}(1+\cos^{2}\theta)^{1/2}=\frac{\cl\rho+2\ce}{\rho(1+\cos^{2}\theta)^{1/2}}>0\,.
\ee
\begin{figure}[pb]
\centerline{\psfig{file=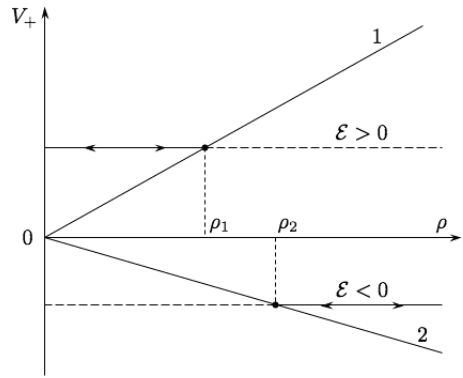,width=6cm}}
\vspace*{8pt}
\caption{A schematic illustration of the effective potential $V_{+}$. Line 1 illustrates positive effective potential. A test particle has positive energy $\ce\geq V_{+}$. Its motion has turning point at $\rho=\rho_{1}$. Line 2 illustrates negative effective potential. If $V_{+}<\ce\leq0$ the particle motion has turning point at $\rho=\rho_{2}$. If the effective potential is zero and energy of the particle is positive, there are no turning points, and if energy of the particle is zero, the particle moves in a marginally stable orbit of $\rho=const\geq0$. \label{f1}}
\end{figure}

\section{Radial motion}

To analyze the radial motion of the particle we consider equation \eq{r}. According to this equation, the radial region of motion is defined by the inequality \eq{c1}
\be\n{rm}
4\ce^{2}+4\ce\cl\rho-\ck\rho^{2}\geq0\,.
\ee
The equality sign corresponds to a point where $\dot{\rho}=0$, which defines a turning point. The relation \eq{rm} implies that either $\ce\leq\ce_{-}$, or $\ce\geq\ce_{+}$, where
\be
\ce_{\pm}=\frac{\rho}{2}\left(-\cl\pm\sqrt{\cl^{2}+\ck}\right)\,.
\ee
This expression is real for $\ck\geq-\cl^{2}$. If this condition does not hold, there are no turning points. However, we may have orbits of constant $\rho$. Such orbits, which are unstable, are defined by $\ck=0,\ce=0$, and $\cl>0$. The local energy of the particle moving in a marginally stable orbit $\rho=const\geq0$ is positive. In the case of turning points, according to the condition \eq{loce}, we must exclude $\ce_{-}$.

Assume now that $\ck\geq-\cl^{2}$, and define an effective potential for the radial region of motion as follows:
\be\n{ep}
V_{+}(\rho):=\ce_{+}\equiv\frac{\rho}{2}\left(-\cl+\sqrt{\cl^{2}+\ck}\right)\,.
\ee
The radial region of motion is possible for $\ce\geq V_{+}(\rho)$. A turning point corresponding to the energy $\ce$ is given by
\be
\rho_{\ce}=\frac{2\ce}{\ck}\left(\cl+\sqrt{\cl^{2}+\ck}\right)\,.
\ee
The local energy \eq{loce} is minimal at the turning point, and one has
\be\n{mloce}
\left.\ce_{{\text{loc}}}\right\vert_{\rho=\rho_\ce}=\frac{\sqrt{\cl^{2}+\ck}}{(1+\cos^{2}\theta)^{1/2}}>0\,.
\ee
Thus, in the case of turning points we must have $\ck> -\cl^{2}$.

The effective potential \eq{ep} is positive if either $\cl<0, \ck>-\cl^{2}$, or $\cl\geq0, \ck>0$; it is zero if $\cl\geq0, \ck=0$, and it is negative if $\cl>0, 0>\ck>-\cl^{2}$. Figure \ref{f1} illustrates the behavior of the effective potential and the radial motion of the particle.

A test particle can have the following types of the radial motion. If the effective potential is positive, energy of the test particle must be positive. In this case the particle always falls down to the horizon, either from the spatial infinity or after it is reflected by the potential barrier at the turning point. If the effective potential vanishes and energy of the test particle is positive, the particle either escapes to the spatial infinity or falls down to the horizon. If energy of the test particle vanishes, the particle moves along a marginally stable orbit of $\rho=const\geq0$. Finally, if the effective potential is negative, the particle has negative energy, it goes to the spatial infinity, directly or after it is reflected by the potential barrier at the turning point. If the particle energy is zero, the particle has formally a turning point at the horizon. If the particle has positive energy, it either escapes to the spatial infinity or falls down to the horizon.

\begin{figure}[pb]
\centerline{\psfig{file=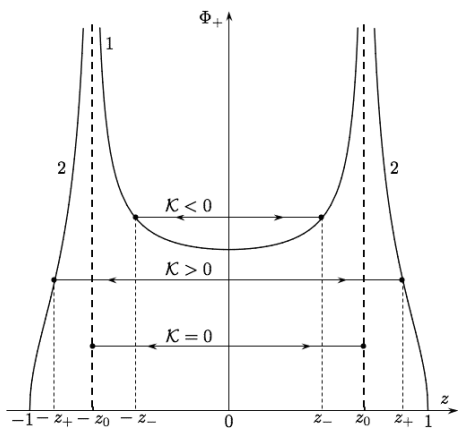,width=6cm}}
\vspace*{8pt}
\caption{A schematic illustration of the effective potential $\Phi_{+}$. Line 1 illustrates the effective potential for $\ck<0$. A photon has angular momentum $\cl\geq \Phi_{+}$. Its motion has turning points at $z=\pm z_{-}$. For $\cl=2\sqrt{|\ck|/3}$ the photon orbit lies in the equatorial plane $z=0$. Line 2 illustrates the effective potential for $\ck>0$. A photon has angular momentum $0\leq\cl\leq \Phi_{+}$. Its motion has turning points at $z=\pm z_{+}$. If $\cl=0$ the photon has turning points at the poles $z=\pm1$. If $\ck=0$, the effective potential is zero. In this case photons which have $\cl>0$, have turning points at $z=\pm z_{0}$, and photons which have $\cl=0$, from the principal null-congruences $z=const$.\label{f2}}
\end{figure}

\section{Polar equations for null rays}

To analyze the polar motion, we consider equation \eq{th}. We first study the case of massless particles $(\varepsilon=0)$. The motion is possible when
\be\n{pmp}
[3-6z^{2}-z^{4}]\cl^{2}+4\ck(1-z^{2})\geq0\,,
\ee
where $z=\cos\theta$. The equality sign corresponds to $\dot{\theta}=0$, which defines a polar turning point. One can see that for $\cl=\ck=0$ there exist orbits with $\theta=const$. Expression \eq{loce} implies that for such orbits $\ce>0$. In this case there are no turning points in the radial region of motion, and the corresponding orbits lie in a cone $\theta=const$. Such orbits define the principal null-congruences of the extreme Kerr throat.
A particular case of this motion is the motion along the axis of rotation $\theta=0$ or $\theta=\pi$.

We can introduce a polar `analogue' of the radial effective potential,
\be\n{epp}
\Phi_{\pm}(z):=\pm2\sqrt{\frac{\ck(1-z^{2})}{z^{4}+6z^{2}-3}}\,.
\ee
A polar turning point occurs at $\cl=\Phi_{\pm}$. The effective potential diverges at $z^{2}_{0}=\sqrt{3}(2-\sqrt{3})\approx0.464$, which  corresponds to $\theta=\theta_{0}\approx0.821$ or $\theta=\pi-\theta_0$. There are two regions: (1) $0\leq|z|\leq |z_{0}|$, where $\ck<0$ and $\cl\leq\cl_{-}$, or $\cl\geq\cl_{+}$, and (2) $|z_{0}|<|z|\leq1$, where $\ck>0$ and $\cl\in[\cl_{-},\cl_{+}]$. The values $\theta_{0}$ and $\pi-\theta_{0}$ define the turning points of the orbits corresponding to $\ck=0$ and $\cl\neq0$, which pass through the equatorial plane $\theta=\pi/2$. The polar motion of a photon is symmetric with respect to the reflection $\cl\to-\cl$. Thus, without loss of the generality, we can take $\cl\geq0$. The behavior of the effective potential for the polar motion of a photon is illustrated in Fig.~\ref{f2}.

The photon has the following types of the polar motion: (1) A motion in the equatorial plane, which corresponds to $\cl=\pm2\sqrt{|\ck|/3}$. This is the minimum of the effective potential with $\ck<0$. Thus, the equatorial polar motion is stable; (2) An oscillatory motion about the equatorial plane with the turning points at $z=\pm z_{-}<z_{0}$ for $\ck<0$, at $z=\pm z_{0}$ for $\ck=0$, and at $z=\pm z_{+}>z_{0}$ for $\ck>0$. In the last case the photon can reach the poles $\theta=0$ and $\theta=\pi$, if $\cl=0$. (3) The photon can move in an orbit confined to a cone $\theta=const$, if $\cl=\ck=0$.

To get a full picture of a photon motion one has to consider the combination of its motion in the radial and in the polar regions\footnote{One needs to include the motion in the azimuthal $\phi$-direction as well. After solving the equations in the $(\rho,\theta)$ sector, one can obtain $\tau(\sigma)$ and $\phi(\sigma)$ by direct integration.}. According to the analysis of the radial motion presented in the previous section, we see that if the photon always falls down to the horizon, its orbit is either confined to the equatorial plane, if $\cl=-2\sqrt{|\ck|/3}$ and $\ck<0$, or it oscillates between the maximal and the minimal values of $\theta$. In particular, if $\cl=0$ and $\ck>0$, it can reach the poles during such oscillations. If the photon can escape to the spatial infinity or to fall down to the horizon (there are no turning points in the radial region of motion), its orbit is confined to a cone $\theta=const$, if $\cl=\ck=0$, or it oscillates between the minimal $\theta=\theta_{0}\approx0.821$ and the maximal $\theta=\pi-\theta_{0}$ values of $\theta$. Additional details of the photon motion are discussed in the caption to Fig.~\ref{f2}.

\section{Polar equations for massive charged particle}

For a massive particle ($\varepsilon=-1$) the following relation must be valid:
\be\n{pmm}
[3-6z^{2}-z^{4}]\cl^{2}+8b(1-z^{4})\cl+4[\ck -1-z^{2}-4b^{2}(1-z^{2})](1-z^{2})\geq0\,.
\ee
The polar equation (\ref{th}) is invariant under the reflection $b\to-b$, $\cl\to-\cl$. Thus, without loss of the generality we can make $b\geq0$.

\begin{figure}[pb]
\centerline{\psfig{file=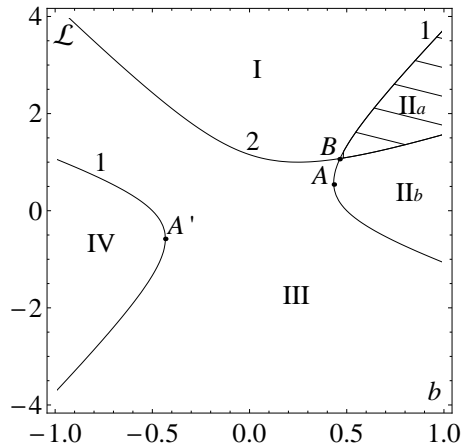,width=6cm}}
\vspace*{8pt}
\caption{Equatorial motion. The left and the right lines 1 are the solutions of the
equation $3\cl^2-8b\cl-16b^2+4=0$. In regions IV and $\text{II}_{a, b}$ the
equatorial motion is unstable. Curve 2 separates two regions of the parameters $\cl$ and
$b$. In region I, above this curve, $V_{+}(\rho)<0$, while in region III, below the curve,
$V_{+}(\rho)>0$. The equatorial instability develops at the critical value of $b=\pm\sqrt{3}/4$
corresponding to $\cl=\pm1/\sqrt{3}$, see points $A$ and $A'$; $\cl=0$ corresponds to
$b=\pm1/2$. Point $B$ is the intersection of the curves 1 (the right branch) and 2, where
$b=3^{1/4}/(2\sqrt{2})$ and $\cl=\sqrt{2}/3^{1/4}$. The region $\text{II}_a$ corresponds
to unstable equatorial motion and infinite motion in the radial direction. \label{f3}}
\end{figure}

Since the problem contains many free parameters, a detailed analysis of different special cases is quite involved. Here we consider only two interesting cases: the motion along the rotation axis, and the equatorial motion. The first
case is trivial. Using equation \eq{th} we find that the motion along the rotation axis $\theta=0$ and $\theta=\pi$ corresponds to $\cl=0$ and $\ck=2$.

Let us now discuss the equatorial motion. Equation \eq{th} can be written as $(z=\cos\theta)$
\be
\dot{z}^{2}=-U(z)\hhh U(z):=\frac{A(z^{4}-1)+2B(z^{2}-1)+4\cl^{2}}{4(1+z^{2})^{2}}\,,
\ee
where $U(z)$ is the angular effective potential. The equatorial plane $\theta=\pi/2$ corresponds to $z=0$. It is a solution of the equation if $U(0)=0$. This gives
\be
3\cl^{2}+8b\cl+4\ck-16b^{2}-4=0\,.\n{st1}
\ee
This solution is stable if the following condition holds:
\be
U''(0)\geq0\,.
\ee
This condition can be written in the form
\be
3\cl^{2}+2\ck-16b^{2}\geq0\,.\n{st2}
\ee
Eliminating $\ck$, we can present the conditions \eq{st1} and \eq{st2} in the $(\cl,b)$-space as follows:
\be
3\cl^{2}-8b\cl-16b^{2}+4\geq0\,.
\ee
When $b=0$, this condition is always satisfied, i.e., the equatorial orbits are stable. For $b\ne0$, only orbits corresponding to sufficiently large values of $|\cl|$ are stable. The regions of stability and instability are shown in Fig.~\ref{f3}.

Let us now consider the combination of the radial and polar motion. The radial effective potential corresponding to the motion along the axis of rotation is
\be
V_{+}(\rho)=\frac{\rho}{\sqrt{2}}\,.
\ee
It is positive in the region outside the horizon $\rho>0$. Thus, energy of a massive charged particle is positive, $\ce>0$, and the particle always falls down to the horizon (see Fig.~\ref{f1}).

Using expressions \eq{ep} and \eq{st1} we derive the radial effective potential corresponding to the equatorial motion
\be\n{repe}
V_{+}(\rho)=\frac{\rho}{4}\left(\sqrt{(\cl-4b)^{2}+4}-2\cl\right)\,.
\ee
The radial effective potential can be positive or negative. If it is positive, a massive charged particle always falls down to the horizon (see Fig.~\ref{f1}). If it is negative, a massive charged particle of negative energy $\ce<0$ always escapes to the spatial infinity. The radial effective potential \eq{repe} is non-positive, if the following condition holds:
\be
3\cl^{2}+8b\cl-16b^{2}-4\geq0\hhh\cl>0\,.
\ee
Curve 2 in Fig.~\ref{f3} illustrates the zero valued effective potential. In the region above the curve the radial effective potential is negative.

\section{Conclusion}

\noindent In this paper, we studied the motion of massless and massive electrically charged particles in a weakly charged extreme Kerr throat. An interesting feature of the extreme Kerr throat is that the Killing tensor of the Kerr spacetime becomes reducible. It means that in the throat the hidden symmetry becomes explicit. We analyzed the radial and the polar equations of motion and demonstrated that the black hole electric charge does not change qualitatively the types of the motion of test particles in the radial direction. However, the charge can have significant effect on the motion of an electrically charged particle in the polar direction. In particular, for $b=qQ/(mM)>\sqrt{3}$ the equatorial motion of the particle may become unstable. This result possibly reflects a simple fact that the electrostatic repulsion becomes comparable to the gravitational attraction. It might be interesting to study this instability when $a/M$ slightly differs from 1.

\section*{Acknowledgments}

The authors wish to thank the Natural Sciences and Engineering Research
Council of Canada for the financial support.  One of the authors
(V.F.) is grateful to the Killam Trust for its support.

\setcounter{footnote}{0}  
\renewcommand{\thefootnote}{\alph{footnote}\alph{footnote}}


\end{document}